\documentclass[graphicx,11pt]{aastex}
\def\icarus{{Icarus}} 
\usepackage{lineno}

\lefthead{} \righthead{}

\begin{document}

 \title{A Sawtooth-like Timeline for the First Billion Years of Lunar Bombardment.} 

\author{\textbf{A. Morbidelli$^{(1)}$, S. Marchi$^{(1,2,3)}$, W.F. Bottke$^{(2)}$, D.A. Kring$^{(3)}$}\\ (1) Dep. Lagrange, UNSA, CNRS, OCA, Nice, France, (Email: morby@oca.eu / Fax: +33-4-92003118), (2) Center for Lunar Origin \& Evolution, SwRI, Boulder,   CO, USA, (3) Center for Lunar Science \& Exploration, USRA - Lunar   and Planetary Institute, Houston, TX, USA} 

\begin{abstract}

We revisit the early evolution of the Moon's bombardment.  Our work combines modeling (based on plausible projectile sources and their dynamical decay rates) with constraints from the lunar crater record, radiometric ages of the youngest lunar basins, and the abundance of highly siderophile elements in the lunar crust and mantle. We deduce that the evolution of the impact flux did not decline exponentially over the first billion years of lunar history, but also there was no prominent and “narrow” impact spike $\sim 3.9$ Gy ago, unlike that typically envisioned in the lunar cataclysm scenario.  Instead, we show the timeline of the lunar bombardment has a sawtooth-like profile, with an uptick in the impact flux near $\sim 4.1$ Gy ago. The impact flux at the beginning of this weaker cataclysm was 5-10 times higher than the immediately preceding period.  The Nectaris basin should have been one of the first basins formed at the sawtooth.  We predict the bombardment rate since $\sim 4.1$~Gy ago declined slowly and adhered relatively close to classic crater chronology models (Neukum and Ivanov (1994)).  Overall we expect that the sawtooth event accounted for about 1/4 of the total bombardment suffered by the Moon since its formation.  Consequently, considering that $\sim 12$-14 basins formed during the sawtooth event, we expect that the net number of basins formed on the Moon was $\sim 45$-50. From our expected bombardment timeline, we derived a new and improved lunar chronology suitable for use on Pre-Nectarian surface units. According to this chronology, a significant portion of the oldest lunar cratered terrains has an age of 4.38-4.42 Gyr. Moreover, the largest lunar basin, South Pole Aitken, is older than 4.3Gy, and therefore was not produced during the lunar cataclysm.

\end{abstract}

\section{Introduction}

The temporal evolution of lunar bombardment is a subject of
intense debate.  A natural expectation is that it declined with time
during the early epochs of solar system history, while planetesimals
left over from planet accretion were in the process of being gradually
removed by dynamical and collisional mechanisms.

In this respect, a surprise came with the first analysis of the lunar
samples collected by the Apollo missions.  They revealed a clustering
of radiometric impact ages at about 3.9 Gy ago (Papanastassiou and
Wasserburg, 1971a, 1971b; Wasserburg and Papanastassiou, 1971; Turner
et al., 1973). Tera et al. (1974) concluded that a major bombardment
episode occurred on the the Moon at that time, i.e. about 0.6 Gyr
after the Moon formation (4.5 Gy ago; see e.g. Kleine et al., 2009),
which they named {\it terminal lunar cataclysm}.  {More recently,
  laboratory analyses on lunar meteorites, which should be more
  representative of the entire lunar surface than the Apollo samples,
  confirmed the strong deficit of impact ages older than $\sim 4$~Gy
  (Cohen et al. 2000), although they did not show a narrow impact
  spike.  This absence of older ages is consistent with the cataclysm
  hypothesis, but it has been argued that it could also be the result
  of biases that work against finding samples with the oldest impact ages
  (Hartmann 1975, 2003; see Hartmann et al., 2000 and Chapman et al.,
  2007 for reviews of the contrasting arguments).}


An analysis of lunar crater densities also fails to yield an
unambiguous view of the temporal evolution of lunar
bombardment. Neukum and Wilhelms (1982) (see also Neukum, 1983; Neukum
and Ivanov, 1994, hereafter NI94) studied the crater density over
terrains of ``known'' radiometric age. {They concluded that the
  bombardment rate was roughly constant (within a factor of 2 or so)
  until 3.5 Gy ago, a result that is generally accepted today. In
  addition, they argued for a long {\it smooth} decay of the impactor
  flux at older times}. Thus, in their model, there is no lunar
cataclysm. The problem, however, is that only the youngest units,
starting with the Imbrium basin ~3.8-3.9 Gy ago, have well established
radiometric ages, whereas the ages of older basins, like Nectaris, are
uncertain (e.g. Norman et al., 2010).  Neukum and collaborators
assumed the age of Nectaris basin was $\sim$4.1 Gy because this age
appears in the samples collected by the Apollo 16 mission that landed
in the lunar highlands near Nectaris (e.g., Maurer et al., 1978). In
this case, the density of craters as a function of age between 4.1 to
3.5 Gy ago seems to decline as $\exp (-a t)$, where $a=6.95$ and $t$
is measured in Gy (see Fig. 1). This exponential evolution was then
extrapolated backwards in time {by NI94}, to estimate the impact
flux during the oldest lunar epochs\footnote{NI94 also used as a
  data-point the density of craters on the lunar highlands, but the
  age of the latter (which NI94 assumed to be 4.35Gy) is not well contrained.}.

A different view is summarized by Ryder (1990), St{\"o}ffler and Ryder
(2001) and Ryder (2002).  They argued that the age of Nectaris is 3.9
Gy because this age appears more prominently than the 4.1 Gy age among
Apollo 16 Descartes terrain samples. If one assumes this age, then the
{\it same} crater counts on Nectaris imply a bombardment rate that has
a {\it much steeper} decline over the 3.5--3.9 Gy period than in NI94.
This steeper decline cannot be extrapolated in time back to the lunar
formation event because it would lead to {unrealistic physical}
implications. For instance, the Moon would have accreted more than a
lunar mass since its formation (Ryder, 2002)!  Consequently this
scenario implies that the bombardment rate could not have declined
smoothly, but rather should have been smaller before 3.9 Gy ago than
in the 3.5-3.9 Gy period, in agreement with the cataclysmic impact
spike hypothesis.  In fact, in an end-member version, Ryder (1990)
also suggested that {\it all} impact basins could have formed during such a
cataclysm impact spike.

The most recent analysis of Apollo 16 samples suggest that the younger
ages from the Descartes terrain are probably ejecta from Imbrium
(Norman et al., 2010). The older ages, however, have no diagnostic
link to Nectaris basin ejecta. This means the age of Nectaris remains
uncertain, and may or may not be represented among Apollo 16
samples. Thus, no definitive conclusion can be derived in favor of the
cataclysm or the smooth exponential decline hypothesis from these
data.

{Other studies on the lunar crater record reported support
  for a lunar cataclysm.  Strom et al. (2005) detected a change in the
  Size Frequency Distribution (SFD) of old craters (i.e. on the
  highlands) relative to young craters (i.e. on the maria plains).
  Marchi et al. (2012) detected the signature of a change in the
  velocity of the projectile populations hitting the Moon at Nectarian
  and pre-Nectarian times respectively. Both findings suggest drastic
  changes in the impactor populations of the solar system, consistent
  with the cataclysm hypothesis.  However, an opposing viewpoint has
  been suggested by Fassett et al. (2012). They also found two
  populations of projectiles, but the transition from one to the other
  occurred in mid-Nectarian epoch, i.e. in the middle of the putative
  cataclysm. They interpreted this result as problematic for the lunar
  cataclysm scenario.  Therefore, it is fair to say that interpreting
  the early cratering record of the Moon is challenging.}

In this paper, we choose not to enter into those technical debates,
but instead revisit the problem with a new combination of theoretical
considerations (by looking at the dynamical evolution of plausible
projectile sources) and existing physical constraints.  More
precisely, we look to calibrate the ``free parameters'' of the problem
(i.e., size of the projectile population, timing of the instability
that released the projectiles from a formerly stable reservoir, the
approximate age of Nectaris basin) to produce a model that is
consistent with (i) the possible dynamical evolution of the solar
system, (ii) the lunar crater record and (iii) particular geochemical
constraints derived from lunar samples. As we will show, our results
support a view that is somewhat intermediate between the two
end-member camps described above: all lunar basins forming in a smooth
decline or a prominent and narrow impact spike 3.9~Gy ago.  In fact,
we will argue for the need of a sudden increase in the lunar impact
rate, but as early as $\sim$4.1-4.2 Gy ago, and not one as pronounced
as in Ryder's (1990) description of the lunar cataclysm. {Our view was
  proposed before (e.g. Fig. 3 in Hartmann et al., 2000), but never
  quantified through a calibrated model.}  Consequently, we believe
the lunar cataclysm implies a decline of the bombardment rate since
4.1 Gy ago, in agreement with that described by NI94.

\section{The Nice model and the E-belt}

The so-called {\it Nice model} (Tsiganis et al., 2005; Morbidelli et
al., 2005; Gomes et al., 2005), named after Nice, France where it was
developed, showed that an impact spike on the terrestrial planets is
possible and plausible due to a sudden change in the orbital
configuration of the giant planets. For a recent review of the
model, the reader can refer to Morbidelli (2010).  For the purposes of
this paper, we limit our discussion to the implications of the model
and how the latest developments affect the lunar cratering record.

The Nice model argues that there were two distinct categories of
projectiles during the impact spike: comets from the trans-Neptunian
disk that likely hit inner solar system targets over a time
span of several tens of millions of years, and asteroids from the
region between Mars and Jupiter, most of which hit over hundreds of
millions of years. Both reservoirs would have been partially
destabilized as the giant planets migrated from their original to
their current orbits.

{The densely cratered surfaces of outer planet satellites like
  Iapetus hint at the possibility that destabilized comets struck the
  Jovian planets' satellites duirng ancient solar system times}, in
agreement with the predictions of the Nice model (Morbidelli et al.,
2005; Nesvorny et al., 2007; Charnoz et al., 2009; Broz et al.,
2011). The evidence for a cometary bombardment becomes more elusive as
one moves toward the inner solar system.  {On the Moon, the SFD of
  the most ancient craters has the same shape as that of main belt
  asteroids (Strom et al., 2005; Marchi et al., 2009, 2012). Also,
  studies of platinum-group elements in ancient lunar samples, which
  presumably were delivered by lunar impactors, show that many
  projectiles were not predominantly composed of primitive,
  carbonaceous chondritic material. This suggests that comets did not
  play a major role in the ancient bombardment (Kring and Cohen, 2002;
  Galenas et al., 2011). The same reasoning can be applied to the
  analysis of the projectile fragments in regolith breccias collected
  at the Apollo 16 site (Joy et al., 2012).  This absence of evidence
  for cometary impactors} can be understood if physical
disintegration, possibly due to explosive ice sublimation, decimated
the cometary population as it penetrated into the inner solar system
(e.g. Sekanina, 1984). The issue is discussed at length in Bottke et
al. (2012).

Concerning asteroids, it was initially thought that objects within the
current boundaries of the asteroid belt would provide a sufficient
source for the lunar cataclysm (Levison et al., 2001;
Gomes et al., 2005). A more detailed study of the orbital evolution of
the terrestrial planets and primordial asteroid belt, however, showed
there is a limit to how much mass could conceivably be extracted
during giant planet migration. Brasser et al. (2009) and Morbidelli et
al. (2010) argued that, among all of the possible giant planet
evolutionary pathways that could take place during the Nice model, the
one that actually occurred had to have been characterized by a fast
displacement of Jupiter's orbit, presumably due to an encounter with
another planet. They named this a {\it jumping-Jupiter-type evolution
  case}\footnote{{The terminology ``jumping-Jupiter'' was
    originally introduced in Marzari and Weidenschilling, 2002, in the
    context of a study on the evolution of extra-solar planets.}}.

Morbidelli et al. (2010) showed that this evolution only removed about
50\% of the asteroids from within the current boundaries of the
asteroid belt, far less than in the non-jumping-Jupiter-type evolution
cases of Levison et al. (2001) and Gomes et al. (2005). An additional
factor of 2 in mass would be lost by main belt objects that suddenly
found themselves within mean motion or secular resonances.  Note that
the interested reader can find a complementary study in Minton and
Malholtra (2011). Together, these works showed that destabilized main
belt asteroids would only produce 2-3 basins on the Moon, not enough
to match lunar cataclysm constraints (Bottke et al., 2012).

The most recent development of the Nice model is the
so-called {\it E-belt} concept (Bottke et al., 2012). It stems from
the realization that the current inner boundary of the asteroid belt
($\sim $2.1 AU) is set by the $\nu_6$ secular resonance whose
existence is specifically related to the current orbits of Jupiter and
Saturn. More specifically, this resonance moves towards the Sun as
the orbital distance between Jupiter and Saturn increases. Moreover,
its strength depends on the eccentricities of the giant
planets. Before the giant planets changed their orbital configuration,
Jupiter and Saturn were closer to one another and were on more
circular orbits; therefore the $\nu_6$ resonance was not present where
it is now: it was located beyond the asteroid belt and it was much
weaker.  Hence the asteroid belt could extend down to the actual
stability boundary set by the presence of Mars (i.e. down to 1.7-1.8
AU, depending of the original eccentricity of the planet). This
putative extended belt population (E-belt) between 1.7-2.1 AU was
almost fully depleted when the orbit of Jupiter (and the $\nu_6$
resonance) ``jumped'' to their current locations.  The few survivors
from the E-belt would now make up the population of Hungaria asteroids
(a group of high-inclined bodies at 1.8-2.0 AU).

There are two free parameters in the E-belt model that need to be set.
One is the total population in the E-belt region. The second one is
the time at which the E-belt was destabilized by the jump of Jupiter's
orbit. Neither are constrained a priori by the dynamical models.

The total E-Belt population was calibrated by Bottke et al. (2012) in
two ways.  The first calibration was provided by the Hungaria
asteroids. Using numerical simulations, they calculated {that
  roughly $\sim 10^{-3}$ of the E-belt population survived in the Hungaria
  region until the present time. Then, using observational constraints
  from the current Hungaria population, they estimated the original
  E-belt population as 1000 times larger.}

The second calibration was provided by the current main belt
population. It is reasonable to expect that the E-belt region was as
densely populated as the rest of the primordial main belt just before
late giant planet migration took place. {Estimating that 75\% of
  the primordial main belt population was removed during resonance
  sweeping (via the jumping Jupiter phase), Bottke et al. used the
  current asteroid population to compute the original orbital density
  of asteroids in the main belt as a function of asteroid size and
  applied it to the E-belt region. The orbital volume of the E-belt is
  about 16--18\% of the main belt orbital volume. Thus, assuming that 
  asteroids had eccentricity and inclination distributions similar to 
 those in the current main belt, this implies that the 
  E-belt carried 16-18\% of the primordial main belt mass, or equivalently 
  60--70\% of the current main belt mass.}

This procedure implicitly assumes that the SFD of the E-belt asteroids
was the same as that of the main belt asteroids, which is reasonable
because the E-belt was simply an extension of the main asteroid
belt. Both SFDs are assumed to be the same as the current SFD of the
main belt, which is justified because the shape of the latter has
probably only experienced minor modifications by collisional evolution
over the last 4 Gy (Bottke et al., 2005; Strom et al., 2005).

The two E-belt calibrations described above yield results similar to
each other, which gives us increased confidence in the coherence of
the E-belt model.

Determining the destabilization time of the E-belt is less
straightforward. The numerical simulations in Bottke et al. (2012)
yield the fraction and impact velocities of the original E-belt
population that should have hit the Moon. From this information, using
the E-belt population described above and applying the crater scaling
relationships described in Schmidt and Housen (1987) and Melosh
(1989), Bottke et al. estimated that the E-belt population should have
caused 9-10 lunar basins (on average). An additional 2-3 basins should
have come from within the current boundaries of the primordial main
belt. Thus, in total, asteroids from the E-belt and main belt would
have caused about 12 basins or so on the Moon. The numerical
simulations suggest that these basin-forming events statistically
should have occurred over a time-span of 400~My, starting from the
E-belt destabilization event\footnote{This timescale is much longer
  than Jupiter's migration timescale because the E-belt objects, after
  being destabilized, can remain for a considerable time on
  meta-stable orbits in the vicinity of the Hungaria region.}. From
the constraint that the youngest lunar basin (Orientale basin) formed
about 3.7--3.8 Gy ago, Bottke et al. deduced that the E-belt
destabilization event occurred $\sim 4.1$--$4.2$~Gy ago. The first of
the basins formed by the E-belt should have occurred very close to
this epoch (see Fig. 4 of Bottke et al., 2012).

\subsection{Comparison with the lunar cratering record}

Using geologic maps and the principle of superposed features, Wilhelms
(1987) deduced that the Nectaris basin was approximately the 12th-14th
youngest basin. Thus, given that the E-belt (and the asteroid belt)
should have produced $\sim 12$ basins, Nectaris is likely to be either
one of the first of the E-belt basins or one of the last basins formed
before the destabilization of the E-belt. However, Marchi et
al. (2012) found evidence that projectiles hitting Nectaris had a
higher impact velocity than those hitting pre-Nectarian terrains such
as the highlands or South Pole Aitken (SPA). This is consistent with
Nectaris being formed within the bombardment caused by the E-belt,
because the latter is characterized by higher impact velocities than
in the previous period (Bottke et al., 2012). These two considerations
together suggest that Nectaris might have been one of the very first
basins formed as a consequence of the destabilization of the
E-belt. To support further this prediction, Bottke et al. (2012)
compared Marchi et al. crater counts on Nectaris terrains to the
expected crater population produced by E-belt objects and found an
excellent match (see Fig. 3 of Bottke et al., 2012).

\begin{figure}[t!] 
\centerline{\includegraphics[height=10.cm,angle=-90]{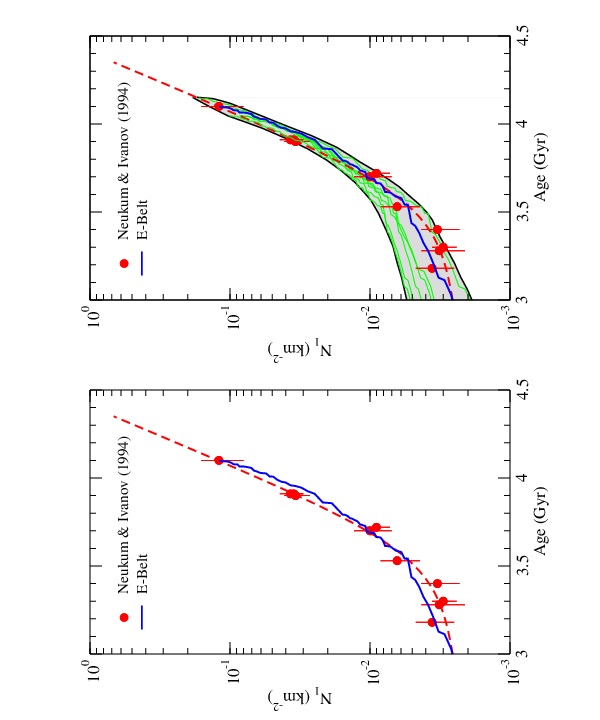}}
\caption{\small In both panels, the dashed red curve shows the density
  of craters larger than 1km diameter ($N_1$) as a function of the
  unit's age, according to NI94. This curve has equation $N_1=
  5.44\times 10^{-14} \left(e^{6.93 t}-1\right)+8.38\times 10^{-4} t$,
  where $t$ is measured in Gy.  The blue curve is the same, but it
  follows the E-belt model (Bottke et al., 2012), assuming that the
  E-belt was destabilized 4.1 Gy ago. The dots with the vertical error
  bars show the density of craters on the terrain calibration units
  according to crater counts and ages reported in NI94. The Nectaris
  basin data-point is the last from the right. Notice that the density
  of $D>1$km craters is not directly measured on the oldest units
  (because of saturation and degradation of small old craters and the
  presence of secondary younger craters); instead, it is estimated
  from the density of larger craters (typically with $D>20$km,
  i.e. $N_{20}$) assuming a SFD for the crater production
  function. Here, for comparative purposes, for the E-belt model we
  have converted $N_{20}$ into $N_1$ using the same crater SFD adopted
  in NI94.  {The green curves in the right panel show different
    E-belt models obtained by changing assumptions on the age of
    Orientale basin and on the total E-belt population within ranges
    that allow to fit the data reasonably well for $t>3.5$~Gy.  We
    also add the contribution of MB-NEAs, assuming different
    bombardment rates, constant with time and up to the current
    value. The shaded area is the envelope of the models that we
    consider acceptable.} }
\end{figure}

Consequently, the Nice/E-belt model predicts that the age of Nectaris
is 4.1--4.2 Gy, in agreement with the assumption of NI94 based on the
radiometric age reported in Maurer et al. (1978), i.e. 4.1~Gy.  Thus,
{in the left panel of Fig. 1 we assume that Nectaris is 4.1 Gy old
  (equivalently, we assume that Orientale is 3.7 Gy old). In this
  case, the curve denoting the density of craters as a function of
  surface age in NI94 and that predicted by the Nice/E-belt model
  match remarkably well over the entire 3.2--4.1 Gy period, a result
  that we did not expect a priori. In fact, the E-belt model was not
  developed to match any specific bombardment timeline, but just to
  complete our understanding of the coupled evolutions of giant
  planets and asteroids.

As explained in the previous section, however, the E-belt model
depends on two parameters: the age of Orientale and the total E-belt
population. The age of Orientale (whose nominal uncertainty is
probably $\sim 100$~My around 3.75 Gy ago) shifts the E-belt cratering
curve along the horizontal axis of Fig.~1. The total E-belt population
(still uncertain by at least a factor of 2 despite of the arguments
based on the Hungaria and main belt populations described above)
shifts the E-belt cratering curve along the vertical axis. In the
right panel of Fig.~1 we have varied these parameters over the range
of values that allow the E-belt cratering curve to fit the data for
$t>3.5$~Gy in an acceptable way. The green curves give examples of the
resulting E-belt cratering curves and the shaded area illustrates the
envelope of the acceptable models. This envelope gives the uncertainty
of ages for a given $N_1$ value. We stress that the models presented
in Fig.~1 also fulfill terrestrial bombardment constraints from impact
spherule beds (Bottke et al., 2012) and not only the lunar cratering
constraints.}

Given the cumulative character of Fig. 1, we have also added the
number of craters generated by E-belt objects alone to those escaping
from the main belt by a combination of Yarkovsky thermal forces and
resonances.  We classify those objects here as main belt-derived
near-Earth asteroids, or MB-NEAs for short. {The crater production
  rate made by MB-NEAs is also uncertain. This uncertainty is included
  as well in the shaded area of Fig.~1, where we assume constant
  cratering rates, from zero up to the current value. The latter is
  probably an upper bound given evidence for an increase in MB-NEA
  flux in the last $\sim 500$~My ago\footnote{Such an increase may be
    due to the formation of the Flora asteroid family (Nesvorny et
    al., 2002, 2007b).}  (Culler et al., 2000; Levine et al., 2005;
  Marchi et al., 2009).}

In summary, the Nice/E-belt model agrees and supports, in broad terms,
the time-line of the lunar bombardment provided by NI94, for times
younger than $\sim 4.1$ Gy ago. {Therefore, in the following, we
  assume the NI94 cratering over this time range, partly because it is
  a standard in the chronology community, but also because we do not
  have a good reason to change it. We now move on to discuss the
  bombardment rate before 4.1 Gy ago.}

\section{The need for a lunar bombardment spike}

The bombardment rate in NI94 before 4.1 Gy has been estimated from a simple backward extrapolation of the bombardment curves calibrated on younger terrains.  Although our model agrees with the NI94 bombardment curve for ages younger than 4.1 Gy, we believe that the extrapolation to older ages is not justified for the following two reasons.

\subsection{Dynamical constraints from inner solar system projectile simulations}

The first reason comes from dynamical considerations, namely that no source of inner solar system projectiles has yet been found that decays over ~1 Gy (say from 4.5 to 3.5 Gy ago) with the rate implied by NI94 curve. For instance, consider the E-belt model, but assume that the destabilization event occurred 4.5 Gy ago. Regardless of the calibration methods discussed in the previous section, assume that the E-belt was about 20 times more populated than in Bottke et al. (2012), so that the cumulative number of craters that it produced on the Moon matches the extrapolation of NI94 curve at 4.5 Gy. As shown in Fig. 2, the model would imply far too many impacts for terrains with ages around 3-4 Gy compared to lunar crater counts. 

\begin{figure}[t!]
\centerline{\includegraphics[height=10.cm,angle=-90]{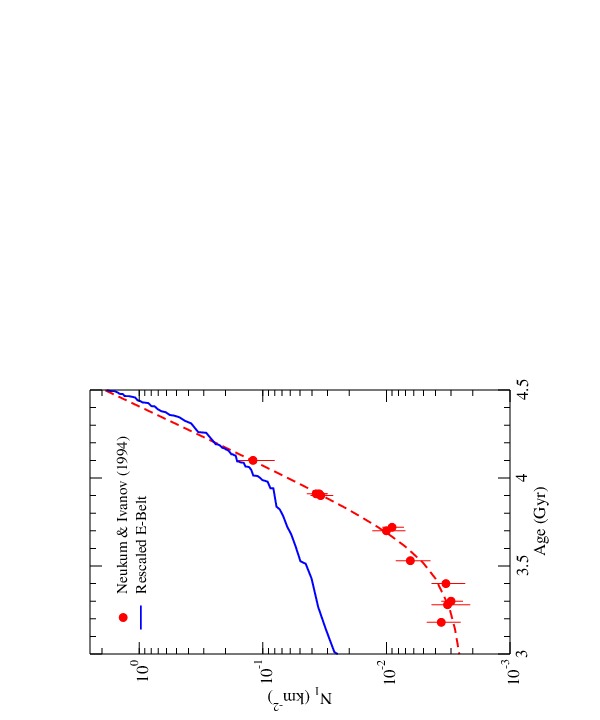}} 
\caption{\small The red curve is the total number of craters larger than 1km diameter  per km$^2$ as a function of unit's age, according to NI94. The dash-blue curve is the same, but here we assumed (i) the E-belt model, (ii) the E-belt was destabilized 4.5 Gy ago and (iii) the E-belt contained 20 times more material than used by Bottke et al. (2012).  This allowed us to match the extrapolation of the NI94 curve at 4.5Gy. Notice the overall mismatch for ages younger than 4 Gy ago.} \end{figure}

We also note that no vertical shift of the E-belt cratering curve in Fig. 2, corresponding to a larger or smaller initial E-belt population, is capable of fitting the NI94 curve at $t\sim 3.5$--4 Gy if the E-belt destabilization event took place 4.5 Gy ago. In fact, the E-belt cratering curve is as steep as NI94 curve near 4.5 Gy, but becomes much shallower at more recent times. This is because most of the bodies surviving for several hundreds of My after the destabilization event are trapped in or near the dynamically sticky Hungaria region. These trapped bodies then leak out from the Hungaria region (developing Earth-crossing orbits) at a very slow rate.  In order for the slopes of the two curves to approximately match one another in the 3.5-4.1 Gy range, the E-belt destabilization event needs to be at $t\sim 4.1$ Gy, as shown in Fig 1.

Once could argue that the E-belt is not the appropriate source of projectiles for the bombardment of the Moon in a scenario without late giant planet migration. The problem, however, is that no appropriate source of projectiles has yet been found using the current system of planets, at least without invoking additional factors to augment the population of the late-arriving projectiles (e.g., the well-timed catastrophic disruption of a Vesta-sized asteroid residing on a Mars- or Earth-crossing orbit 3.9 Gy ago; Cuk 2012).  In fact, Bottke et al. (2007) made a general argument against the possibility that such a source could exist. 

Consider that all comprehensive lunar bombardment models need to produce the 900 and 1200 km diameter basins Imbrium and Orientale between 3.85 and 3.7 Gy ago, and no further basin formation events since that time. This requires a relatively fast decaying impacting population at $\sim 3.8$Gy (unlike the E-belt example above that destabilized 4.5 Gy ago).  Moreover, the decay would have to have been even faster earlier on, because population decay rates typically slow down with time.  Thus, the original population would have been implausibly large, of the order of a few Earth masses of material. It is unlikely that such a population existed at the end of terrestrial planet formation, otherwise the terrestrial planets would have grown more massive.

\subsection {Geochemical constraints from the Moon} \label{2r}

The second reason for not believing the extrapolation of the NI94
curve before 4.1 Gy is provided by lunar geochemical constraints. It
can be argued that the Moon's formation in a giant impact on Earth and
its subsequent differentiation and magma ocean phase should have sent
most iron and highly siderophile elements (HSEs) to the core. Although
we have no direct samples of lunar mantle rocks, studies of HSE 
isotopes in derivative lunar mantle melts suggest that the Moon
accreted no more than $1.7\times 10^{19}$kg of chondritic material
before its mantle became protected from incoming material by a thick
crust (Walker et al., 2004; Day et al. 2007, 2010; see also Bottke et
al. 2010). The lunar crust, however, is also not very rich in HSE: the
amount of chondritic material delivered into the lunar crust after its
formation is estimated to be $0.4\times 10^{19}$~kg (Ryder,
2002). Thus, in total, the Moon should have accreted less than $\sim
2.1\times 10^{19}$~kg of material. Artemieva and Shuvalov (2008)
estimated that, for asteroids with an impact velocity of 18km/s (which
is intermediate between the velocities of projectiles hitting the Moon
before and after the destabilization of the E-belt; see
sect.~\ref{preLHB}) and an impact angle of $45^\circ$, only 60\% of
the impactor's material is effectively accreted, the rest being lost
into space (e.g., see also Bottke et al. 2010).  This implies that the
total mass of impactors hitting the Moon since its formation was at
most $\sim 3.5\times 10^{19}$~kg.

{The abundance of HSE is remarkably similar in enstatite, ordinary
  and charbonaceous chondrites (see Table 1 in Walker, 2009). Thus the
  constraints on the accreted mass reported above are presumably valid
  for broadly chondritic projectiles striking the Moon. However, if
  the bombardment had been dominated by objects derived from the
  mantles of differentiated bodies, the amount of HSEs delivered per
  unit mass would have been much smaller. In other words, the mass
  accreted by the Moon after its formation could have been higher than
  estimated above.  We consider that this possibility is unlikely
  because: (i) achondritic meteorites are rare and (ii) the SFD of the
  craters on the oldest lunar terrains suggests that the projectiles
  had a SFD similar to main belt asteroids (Strom et al., 2005), which
  would be surprising if the objects had had predominantly a different
  physical nature; (iii) the HSEs in the Moon and terrestrial mantle
  are in chondritic proportions, whereas they would be fractionated
  relatively to each other if they had been delivered predominantly by
  achondritic objects.  For these reasons, it is standard practice to
  use HSE abundances as indicators of the accreted massed assuming
  chondritic proportion (e.g.  Ryder 1990; Walker 2009; Day et
  al. 2007, 2010).}

Below, we compare this value to the total mass of
projectiles that should have hit the Moon in the bombardment history
of NI94.  This estimate is done in two steps.

\noindent{\it Step I:} First, we extrapolated the NI94 curve to 4.5~Gy, the approximate time of the Moon's formation (Kleine et al., 2009).  This yielded the number of craters larger than 1 km diameter per surface square kilometer $N_1=2.7$. The relationship between $N_1$ in NI94 and $N_{20}$ in Marchi et al. (2012) (respectively, the number of craters larger than 1 and 20 km per surface square kilometer) is $\sim$1400. Assuming this ratio, the NI94 curve, extrapolated to 4.5 Gy implies $N_{20}=1.9\times 10^{-3}$. 

\noindent{\it Step II:} Next, we established a general procedure to link $N_{20}$ to the total mass of the corresponding projectile population. The procedure is as follows.  First, we computed the impactor size $D_{20}$ corresponding to a crater of 20~km.  For this, we use the correction from final-to-transient crater as reported in Marchi et al. (2011) and we adopt the Pi-scaling group law for hard rock in the formulation of Schmidt and Housen (1987) and Melosh (1989), assuming an impact velocity of 18 km/s.   Second, we assume that the projectile SFD is the same as the main belt SFD (Strom et al., 2005). Given the numbers of projectiles larger than $D_{20}$ (which is the product between the $N_{20}$ value and  the surface of the Moon in km$^2$), we estimate the size $D_{\rm max}$ of the largest projectile that should have impacted the lunar surface.  Given this information we finally compute the total projectile mass. For this, we assumed a projectile density of $\rho=2.6$ g/cm$^3$, a scaled value designed to account for the mass integral of the entire main belt SFD over all compositions, with the estimated total mass of the asteroid belt given by Krasinsky et al. (2002). Note that because $D_{\rm max}$ changes with $N_{20}$, the total mass of the projectiles hitting the Moon does not scale linearly with $N_{20}$.

Applying these two steps, we estimate that the total mass of the projectiles hitting the Moon since its formation in the NI94 bombardment history is $1.3\times 10^{20}$~kg. This is a factor of 4 larger than the upper bound on the total mass delivered to the Moon throughout its history, as constrained above from the lunar HSE abundances.

\subsection{Summary}

The two reasons discussed above suggest the need for a break, or inflection point, in the bombardment curve at sometime in the 4.1-4.2 Gy interval. While the bombardment rate just before the break had to be smaller than after the break, the impact flux probably increased from that point backwards in time until the Moon-formation event. This discontinuity defines the signature of a lunar cataclysm.

In the next section we attempt to derive a plausible lunar impact rate in the 4.1--4.5 Gy period using both dynamical considerations and the constraints provided by the abundance of lunar HSEs.

\section{The nature of the lunar bombardment before the cataclysm}
\label{preLHB}

The bombardment of the terrestrial planets in the early epochs of the solar system, well before the lunar cataclysm, was presumably caused by remnant planetesimals from the original disk that formed the planets. The best available computer simulations of the terrestrial planet accretion process are those reported in Hansen (2009) and Walsh et al. (2011), mainly because they can satisfactorily reproduce the mass distribution and orbital characteristics of the terrestrial planets.

In order to understand the dynamics of planetesimals leftover from the planet accretion process, we considered 4 of the most successful simulations from Walsh et al. (2011). In two of the simulations, the terrestrial planets reached completion and stabilization in $\sim 30$~My (simulations A). In the other two they stabilized at $\sim 50$~My (simulations B). This is acceptable because the time required for the formation of the Earth is only modestly constrained by radioactive chronometers (see Kleine et al., 2009, for a review).  A timescale of 30 to 50 My is considered a realistic timescale, although $\sim 100$~My has also been suggested (e.g. Allegre et al., 1995; Touboul et al., 2007). The simulations of Hansen and Walsh et al., however, always complete the formation of the terrestrial planets well before the latter time.

For the simulations A and B, we took the orbital distribution of the planetesimals surviving at 30 or 50 My, respectively. We cloned the planetesimals by randomizing the orbital angles (mean anomaly, longitude of node and of perihelion).  This gave us  4 sets with a total of 2,000 particles each.

The final synthetic terrestrial planets in the Walsh et al. simulations form a system relatively similar but not identical to our own. Thus, to study the dynamical decay of the planetesimal populations in the actual solar system, we need to substitute the synthetic planets with the ``real'' ones.

The problem is that eccentricities and inclinations of the terrestrial
planets before giant planet migration (at the time of the lunar
cataclysm) are uncertain. Brasser et al (2009) argued that the orbits
of the terrestrial planets might have been significantly more circular
and less inclined than the current orbits. They could not exclude the
possibility, however, that the terrestrial planets' eccentricities and
inclinations were already comparable to the current ones.  Thus, for
each of the 4 sets of planetesimals we did two integrations. For the
first, we assumed that the terrestrial planets had their current
orbits.  For the second, we put the terrestrial planets on orbits with
their current semimajor axes but with eccentricities and inclinations
equal to zero.

Our integrations covered 400 My (i.e. the time-span between the Moon
forming event and the onset of the lunar cataclysm, assuming that they
happened respectively 4.5 and 4.1 Gy ago).  At each output time, we
then computed the collision probability and impact velocity of each
particle with the Earth ($c_p(t)$) using the algorithm described in
Wetherill (1967): the semimajor axis, eccentricity and inclination of
the particle and the Earth were kept equal to the values registered in
the output, while the angles (mean anomaly, longitude of perihelion
and of the node) were randomized over 360 degrees.  The effect of the
Earth's gravitational focusing was also taken into account, {given
  the relative velocities provided by the simulations}. The impact
probabilities of all particles at a given time were then summed,
obtaining a total collision probability ($C_E(t)=\sum_p c_p(t)$) at
the considered time.  The collision probability with the Moon $C_M(t)$
was assumed to be a constant fraction (1/20) of $C_E(t)$. The actual
value of this ratio (which depends on the velocity of the projectiles)
is not important, as we are only interested here in the time evolution
of $C_M$ and not its absolute value. In principle the $C_M/C_E$ ratio
decreases with time as the Moon gets farther from the Earth following
its tidal evolution. But in practice Moon's migration is very fast at
the very beginning and then slows down considerably, so that the
assumption that $C_M/C_E$ is constant is a classical, reasonable
approximation.

The tabulated function $C_M(t)$ describes the time evolution of the lunar impact rate in the considered simulation. It was then interpolated with a function of type $\exp(-(t/t_0)^\beta)$ (Dobrovolskis et al., 2007) to obtain a smooth, analytical function. The decay is obviously different from simulation to simulation, but we found it to be confined between two functions with ($t_0=10$My, $\beta=0.5$) and ($t_0=3$My, $\beta=0.34$). At 400~My, the values of these two functions have a ratio of 2.5.  

Recall that the total mass accreted by the Moon since it
differentiated, according to constraints from lunar HSEs, is $\lesssim
3.5\times 10^{19}$~kg. Given the value of $N_{20}$ for the Nectaris
basin ($8.66 10^{-5}$km$^{-2}$: Marchi et al., 2012) and applying the
procedure explained in sect.~\ref{2r}, we find that the net projectile
mass that has hit the Moon since 4.1 Gy ago is $2\times 10^{18}$~kg.
By subtracting this value from $\sim 3.5\times 10^{19}$~kg we conclude
that $\lesssim 3.3\times 10^{19}$~kg of projectiles should have hit
the Moon between 4.1 and 4.5 Gy ago.

Taking the functions that bracket the decay of the impact rate, we normalized both of them so that the total mass hitting the Moon in the 4.1-4.5 Gy period was $3.3\times 10^{19}$~kg. The correspondence between $N_{20}$ and the mass hitting the Moon was computed using the procedure described in sect.~\ref{2r}\footnote{In agreement with the results of Marchi et   al. (2012), we used an impact velocity of 11 km/s instead of 18 km/s as stated in sect.~\ref{2r} for the computation of the projectile size needed to make a 20 km crater.}.

 \begin{figure}[t!] 
\centerline{\includegraphics[height=10.cm,angle=-90]{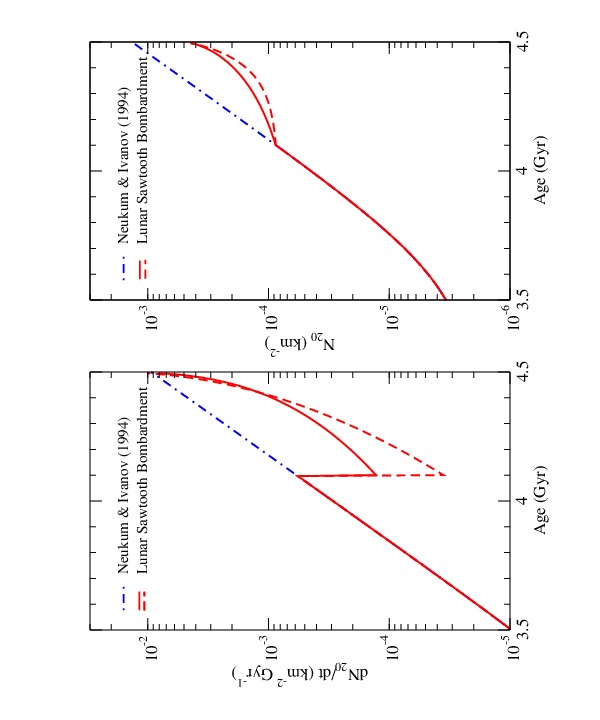}} 
\caption{\small The timeline of lunar bombardment. The blue
  dash-dotted line is the exponential decay assumed by NI94.  The red
  curve are our results combining a post-accretion bombardment of
  leftover planetesimals in the 4.1--4.5 Gy time range and the
  asteroid main belt and E-belt bombardments for $t<4.1$Gy. The solid
  and dashed curves for $t>4.1$ Gy correspond to the two analytic
  functions that bracket the early decay of the cratering rate, as
  discussed in the text.  The left panel shows the bombardment rate as
  a function of time (measured in number of craters larger than 20 km
  produced per km$^2$ and per Gy).  The right panel shows the
  cumulative bombardment suffered by a given terrain as a function of
  its age (measured in number of craters larger than 20km per km$^2$,
  i.e. $N_{20}$). In the left panel, the equations of for the curves
  are the following. For $t<4.1$Gy: d$N_{20}$/d$t=2.7\times
  10^{-16}e^{6.93 t}+5.9\times 10^{-7}$; for $t>4.1$Gy and the solid
  curve: d$N_{20}$/d$t= 2.5\times 10^{-2}e^{-[(4.5-t)/0.003]^{0.34}}$;
  for the red dashed curve curve: d$N_{20}$/d$t= 2.0\times
  10^{-2}e^{-[(4.5-t)/0.01]^{0.5}}$; the time $t$ is measured in Gy.}
\end{figure}

Fig. 3 summarizes our results. The red curves on the left panel shows our estimate of the time evolution of the lunar impact rate.  These curves have a {\it sawtooth} profile, with the uptick at 4.1~Gy representing the onset of the lunar cataclysm. This profile is very different than the continuous exponential decay of NI94 for ages older than 4.1 Gy, as shown by the blue line. The right panel shows the cumulative lunar impact flux as measured by the density of craters expected on a given surface as a function of its age. The curves were obtained by integrating the values reported on the left panel from 0 to $t$ for all values of $t$.

\section{Discussion and Conclusions}

{We have taken our best models of the early solar system
  evolution, determined the impact flux on the Moon over time and then
  calibrated these results using the existing dynamical, geochemical,
  and crater density constraints.

We infer that the evolution of the lunar bombardment rate}
is somewhat intermediate between the two end-member views in this
historical controversy.  Our model bombardment rate from 4.1 to 3.5 Gy
ago agrees with the exponential decay illustrated by NI94, the
champions of the no-cataclysm view. We find that it is impossible,
however, to extrapolate their exponential flux backward in time before
4.1 or possibly 4.2 Gy. We believe a discontinuity in the evolution of
the bombardment rate, or a lunar cataclysm, is the easiest way to
match constraints.  The timeline of the Moon's bombardment that
emerges from our study has a sawtooth profile, with a moderate uptick
at 4.1--4.2 Gy (see Fig 3, left panel).  This stands in sharp contrast
with the prominent impact spike usually shown in sketches of the lunar
cataclysm; {instead it is in broad agreement with the scenario of
  ``weak cataclysm'' promoted in Fig. 3 of Hartmann et al. (2000).}

Our impact flux model predicts the lunar bombardment from 4.1 Gy ago
--which includes the cataclysm caused by the destabilization of the
main asteroid belt ($\sim 15-25$\% of cataclysm impactors) and E-belt
($\sim 75-85$\% of cataclysm impactors)-- accounts for approximately
25\% of the total bombardment suffered by the Moon since it
formed. This is in agreement with the total number of basins on the
Moon ($\sim 50$), of which only $\sim 12$-14 are Nectarian and
early-Imbrian (i.e. younger than $\sim 4.1$ Gy).  We note that we
could be underestimating the number of basins because some ancient
basins have probably been erased (Frey and Romine, 2011). The best
estimate at present is that the total number of basins probably does
not exceed $\sim 60$ (Frey, private communication), not far from our
expectations for the combined leftover planetesimal and E-belt models.
We stress that basin formation is necessarily a stochastic process
which can deviate from probabilistic expectations due to small number
statistics.  {Assuming Poisson statistics (i.e. the error bar on
  $N$ events is $\sqrt{N}$), the number of basins formed in our model
  since the destabilization of the E-belt is $12 \pm 4$, while the
  total number of basins is $45 \pm 7$.}

Large portions of the lunar highlands have a crater density that is
about twice that of Nectaris (Strom, 1977; Marchi et al.,
2012), with a value of $N_{20}=1.73\times 10^-4$. 
According to the cumulative bombardment shown in Fig. 3, this
would imply that these portions of highlands started to retain craters
about 4.38-4.42 Gy ago, ages that are consistent with recent estimates
of the timescale for the thickening of the lunar lithosphere (Meyer et
al. 2010). This age also approximates the closure age of the crust, as
derived from zircons that crystallized in the remaining urKREEP
residue of the lunar magma ocean 4.38 to 4.48 Gy (Nemchin et al. 2009
; Taylor et al. 2009). {Similarly, the value of $N_{20}$ for the SPA
floor is $1.36\times 10^{-4}$ (Marchi et al., 2012). This implies
that craters started to accumulate on SPA since 4.33-4.39~Gy ago.
This age should be considered as a lower bound for the formation age of
SPA, because the basin's floor might have solidified only after
some time; it clearly shows that SPA is an old basin, which definitely
predates the cataclysm event.}

The sawtooth-like bombardment timeline has important implications for Earth's habitability.  In the no-cataclysm view, the Earth was increasingly hostile to life going back in time, as the bombardment exponentially increased.  In the classic view of the lunar cataclysm, the prominent impact spike 3.8-3.9 Gy ago conceivably sterilized the Earth by vaporizing all the oceans and thereby creating a steam atmosphere (Maher and Stevenson, 1988; however see Abramov and Mojzsis, 2009). In our sawtooth view, big impactors hit over an extended period, with more lulls and therefore more opportunities for the Hadean-era biosphere to recover. Perhaps in this scenario, life formed very early and has survived in one form or another through the lunar cataclysm.

\vskip 20pt
{\bf Acknowledgments}

We thank R. Walker, W. Hartmann and an anonymous reviewer for the
comments provided to the first version of this manuscript.
A. Morbidelli and S. Marchi thank Germany’s Helmholtz Alliance for
providing support through their “Planetary Evolution and Life”
programme.  The contributions of S.Marchi, W. F. Bottke, and
D. A. Kring were supported by NASA's Lunar Science Institute (Center
for Lunar Origin and Evolution at the Southwest Research Institute in
Boulder, CO; Center for Lunar Science and Exploration at the Lunar and
Planetary Institute in Houston, TX).


\section{References}

\begin{itemize}

\item[-]  Abramov, O., \& Mojzsis, S.~J.,\ 2009.\ Microbial habitability of the Hadean Earth during the late heavy bombardment.\ Nature 459, 419-422. 
\item[-]  All{\`e}gre, C.~J.,  Manh{\`e}s, G., \& Gopel, C.,\ 1995, \gca, 59, 2445  
\item[-] Artemieva N.~A., \& Shuvalov V.~V., 2008, SoSyR, 42, 329  
\item[-] Bottke, W.~F., Durda,  D.~D., Nesvorn{\'y}, D., et al.,\ 2005, \icarus, 175, 111   
\item[-] Bottke, W.~F., Levison,  H.~F., Nesvorn{\'y}, D., \& Dones, L.,\ 2007, \icarus, 190, 203  
\item[-] Bottke W.~F., Vokrouhlick{\'y} D., Minton D., Nesvorn{\'y} D., Morbidelli A., Brasser R., Simonson B., \& Levison H.~F., 2012, Nature, 485, 78 
\item[-] Brasser, R., Morbidelli, A., Gomes, R., Tsiganis, K., \& Levison, H.~F.,\ 2009, \aap, 507, 1053 
\item[-] Bro{\v z}, M.,  Vokrouhlick{\'y}, D., Morbidelli, A., Nesvorn{\'y}, D.,  \& Bottke, W.~F.,\ 2011, \mnras, 414, 2716  
\item[-] Chapman, C.~R., Cohen,  B.~A., \& Grinspoon, D.~H.,\ 2007, \icarus, 189, 233 
\item[-] Charnoz, S.,  Morbidelli, A., Dones, L., \& Salmon, J.,\ 2009, \icarus, 199, 413  
\item[-] {\'C}uk, M.,\ 2012.\ Chronology and sources of lunar impact bombardment.\ Icarus 218, 69-79.
\item[-] Culler T.~S., Becker T.~A., Muller R.~A., \& Renne P.~R., 2000, Sci, 287, 1785  
\item[-] Day J.~M.~D., Pearson D.~G., \& Taylor L.~A., 2007, Sci, 315, 217 
\item[-] Day J.~M.~D., Walker R.~J., James O.~B., \& Puchtel I.~S., 2010, E\&PSL, 289, 595 
\item[-]  Dobrovolskis,  A.~R., Alvarellos, J.~L., \& Lissauer, J.~J.,\ 2007, \icarus, 188, 481  
\item[-] Fassett, C.I. et al., 2012, Earth and Planetary Science Letters, in press. 
\item[-] Frey, H.~V., \& Romine, G.~C.,\ 2011, Lunar and Planetary Institute Science Conference Abstracts, 42, 1190 
\item[-] Galenas, M.~G.,  Gerasimenko, I., James, O.~B., Puchtel, I.~S.,  \& Walker, R.~J.,\ 2011, Lunar and Planetary Institute Science Conference Abstracts, 42, 1413  
\item[-] Gomes, R., Levison,  H.~F., Tsiganis, K., \& Morbidelli, A.,\ 2005, \nat, 435, 466   
\item[-] Hartmann W.~K., 1975, Icar, 24, 181
\item[-] Hartmann W.~K., 1975, Icar, 24, 181 
\item[-] Hartmann, W.~K.,  Ryder, G., Dones, L.,  \& Grinspoon, D.,\ 2000, Origin of the Earth and Moon, 493 
\item[-] Hansen, B.~M.~S.,\ 2009, \apj,  703, 1131  
\item[-] Housen, K.~R., Schmidt,  R.~M., \& Holsapple, K.~A.,\ 1983, \jgr, 88, 2485  
\item[-] Joy K.~H., Zolensky M.~E., Nagashima K., Huss G.~R., Ross D.~K., McKay 
D.~S., \& Kring D.~A., 2012, Sci, 336, 1426 
\item[-] Kleine, T., Touboul, M.,  Bourdon, B., et al.,\ 2009, \gca, 73, 5150 
\item[-]  Kring, D.~A., \& Cohen, B.~A.,\ 2002, Journal of Geophysical Research (Planets), 107, 5009  
\item[-] Levine J., Becker T.~A., Muller R.~A., \& Renne P.~R., 2005, GeoRL, 32, 15201 
\item[-]  Levison, H.~F., Dones,  L., Chapman, C.~R., et al.,\ 2001, \icarus, 151, 286  
\item[-]  Maher, K.~A., \& Stevenson, D.~J.,\ 1988.\ Impact frustration of the origin of life.\ Nature  331, 612-614. 
\item[-]  Marchi, S., Mottola, S.,  Cremonese, G., Massironi, M., \& Martellato, E.,\ 2009, \aj, 137, 4936  
\item[-] Marchi S., Massironi M., Cremonese G., Martellato E., Giacomini L., Prockter L., 2011, P\&SS, 59, 1968  
\item[-] Marchi, S., Bottke, W.F., Kring, D.A., \& Morbidelli, A., 2012, E\&PSL, 325, 27  
\item[-] Marzari F., \& Weidenschilling S.~J., 2002, Icar, 156, 570 
\item[-] Maurer, P., Eberhardt,  P., Geiss, J., et al.,\ 1978, \gca, 42, 1687  
\item[-] Melosh H.J,. 1989. Impact cratering: A geologic process. Oxford University Press. 
\item[-]  Meyer, J., Elkins-Tanton,  L., \& Wisdom, J.,\ 2010, \icarus, 208, 1  
\item[-] Minton, D.~A., \& 
Malhotra, R.,\ 2011.\ Secular Resonance Sweeping of the Main Asteroid Belt 
During Planet Migration.\ The Astrophysical Journal 732, 53. 
\item[-]  Morbidelli, A.,  Levison, H.~F., Tsiganis, K., \& Gomes, R.\, 2005, \nat, 435, 462  
\item[-] Morbidelli, A., 2010  Comptes Rendus Physique, 11, 651-659 
\item[-] Morbidelli, A.,  Brasser, R., Gomes, R., Levison, H.~F.,  \& Tsiganis, K.,\ 2010, \aj, 140, 1391  
\item[-] Nemchin, A.~A., Pidgeon, R.~T., Healy, D., Grange, M.~L., Whitehouse, M.~J., \& Vaughan, J.,\ 2009. M\&PS 44, 1717-1734. 
\item[-]Nesvorn{\'y} D., Morbidelli A., 
Vokrouhlick{\'y} D., Bottke W.~F., \& Bro{\v z} M., 2002, Icar, 157, 155 
\item[-] Nesvorn{\'y}, D.,  Vokrouhlick{\'y}, D., \& Morbidelli, A.,\ 2007, \aj, 133, 1962      
\item[-] Nesvorn{\'y}, D., Vokrouhlick{\'y}, D., 
Bottke, W.~F., Gladman, B., \& H{\"a}ggstr{\"o}m, T., 2007b, Icar, 188, 400 
\item[-] Neukum, G., \& Wilhelms, D.~E.,\ 1982, Lunar and Planetary Institute Science Conference Abstracts, 13, 590  
\item[-] Neukum, G.,\ 1983, Habilitation disseration for faculty membership, Univ. of Munich. 
\item[-] Neukum, G. \& Ivanov, B.A., 1994, in Hazards due to Comets and Asteroids, p. 359. 
\item[-]  Norman, M.~D., Duncan,  R.~A., \& Huard, J.~J.,\ 2010, \gca, 74, 763  
\item[-] Papanastassiou, D.~A., \& Wasserburg, G.~J.,\ 1971a, Earth and Planetary Science Letters, 12, 36  
\item[-] Papanastassiou, D.~A., \& Wasserburg, G.~J.,\ 1971b, Earth and Planetary Science Letters, 11, 37  
\item[-] Ryder, G.,\ 1990, Eos, 71, 313
\item[-] Ryder, G.,\ 2002, Journal of  Geophysical Research (Planets), 107, 5022  
\item[-] Schmidt R.~M., \& Housen K.~R., 1987, IJIE, 5, 543 
\item[-] Sekanina, Z.,\ 1984, \icarus,  58, 81  
\item[-] St{\"o}ffler, D., \& Ryder, G.,\ 2001, \ssr, 96, 9  
\item[-] Strom, R.~G., 1977. Phys. Earth Planet. Inter. 15, 156. 
\item[-] Strom, R.~G., Malhotra,  R., Ito, T., Yoshida, F., \& Kring, D.~A.,\ 2005, Science, 309, 1847  
\item[-] Taylor D.~J., McKeegan K.~D., Harrison T.~M., \& Young E.~D., 2009, GeCAS, 73, 1317 
\item[-] Tera, F., Papanastassiou, D.~A., \& Wasserburg, G.~J.,\ 1974, Earth and Planetary Science Letters, 22, 1  
\item[-] Touboul, M., Kleine,  T., Bourdon, B., Palme, H., \& Wieler, R.,\ 2007, \nat, 450, 1206  
\item[-] Tsiganis, K., Gomes,  R., Morbidelli, A., \& Levison, H.~F.,\ 2005, \nat, 435, 459  
\item[-] Turner, G., Cadogan, P.~H., \& Yonge, C.~J.,\ 1973, \nat, 242, 513  
\item[-] Walker R.~J., Horan M.~F., Shearer C.~K., \& Papike J.~J., 2004, E\&PSL, 224, 399 
\item[-] Walker, 
R.~J., 2009, ChEG, 69, 101 
\item[-] Walsh, K.~J., Morbidelli,  A., Raymond, S.~N., O'Brien, D.~P., \& Mandell, A.~M.,\ 2011, \nat, 475, 206  
\item[-] Wasserburg, G.~J., \& Papanastassiou, D.~A.,\ 1971, Earth and Planetary Science Letters, 13, 97  
\item[-] Wetherill, G.~W.,\ 1967,  \jgr, 72, 2429  
\item[-] Wilhelms, D.E., 1987. The geologic history of the Moon. (U.S. Geol. Surv. Prof. Pap., 1348).

\end{itemize}

\end{document}